\documentclass{Interspeech2024}
\usepackage{multirow}
\usepackage{cite}

\interspeechcameraready

\title{MR-RawNet: Speaker verification system with multiple temporal resolutions for variable duration utterances using raw waveforms}

\name[]{Seung-bin}{Kim}
\name[]{Chan-yeong}{Lim}
\name[]{Jungwoo}{Heo}
\name[]{Ju-ho}{Kim}
\name[]{Hyun-seo}{Shin}
\name[]{Kyo-Won}{Koo}
\name[]{Ha-Jin}{Yu$^\dagger$}
\address{University of Seoul, Republic of Korea\thanks{$^\dagger$Corresponding author.}}

\email{kimho1wq@naver.com, cksdud585@naver.com, jungwoo4021@gmail.com, wngh1187@naver.com,  gustjtls123@naver.com, kkwr0504@naver.com, hjyu@uos.ac.kr}

\keywords{speaker verification, raw waveform, short duration, multi resolution}

\begin{document}

\maketitle

\begin{abstract}
In speaker verification systems, the utilization of short utterances presents a persistent challenge, leading to performance degradation primarily due to insufficient phonetic information to characterize the speakers.
To overcome this obstacle, we propose a novel structure, MR-RawNet, designed to enhance the robustness of speaker verification systems against variable duration utterances using raw waveforms. 
The MR-RawNet extracts time-frequency representations from raw waveforms via a multi-resolution feature extractor that optimally adjusts both temporal and spectral resolutions simultaneously.
Furthermore, we apply a multi-resolution attention block that focuses on diverse and extensive temporal contexts, ensuring robustness against changes in utterance length.
The experimental results, conducted on VoxCeleb1 dataset, demonstrate that the MR-RawNet exhibits superior performance in handling utterances of variable duration compared to other raw waveform-based systems.

\end{abstract}

\vspace{-0.15cm}
\section{Introduction}

Speaker verification (SV) is the task of verifying whether an anonymous speaker is the target speaker registered in the system.
With the development of deep neural networks (DNNs), traditional machine learning-based SV systems have been largely replaced by DNN-based SV systems \cite{dehak2011front, variani2014deep, snyder2018x}.
Despite the exceptional performance achieved by DNN-based SV systems, most systems are typically evaluated using long utterances \cite{desplan2019ecapa, mfaconformer, nextdtnn}.
However, in real environments, shorter utterances of 1 to 2 seconds are often encountered, and SV systems experience a performance degradation as the length of the utterance decreases.
This degradation occurs because short utterances may not contain sufficient speaker-specific phonetic characteristics that can be obtained from speech \cite{kim2020seagg, hajavi2019short, rawnext}.
Thus, SV systems should be constructed to yield similar performance across utterances of various lengths, particularly shorter ones.

To enhance robustness against variable lengths, we initially focused on the input features of the system.
The inputs for SV systems can be traditional handcrafted features such as MFCCs or Mel-filter banks \cite{variani2014deep, snyder2018x, desplan2019ecapa, mfaconformer}, or raw waveforms, which have recently begun to be utilized \cite{jung2017rawcnn, CLDNN, rawnet2, rawnet3}. 
While handcraft features can provide refined information through processing based on human knowledge, they may also offer information loss compared to the unprocessed raw waveforms \cite{wav2vec, wav2vec2.0, hubert}. %\cite{wavlm}
Especially for SV systems that consider limited information, as much information as possible should be available from the input.
Therefore, we use raw waveforms as input to ensure robustness to utterances of various lengths, which have high potential due to the absence of information loss.

In traditional raw waveform-based systems, a one-dimensional input is converted into a two-dimensional feature map through a special module in the first layer, such as \cite{sincnet, pari2020filter}.
The first-layer module extracts meaningful representations from the input, typically based on a fixed frame size.
However, representations based on a constant frame size are unable to offer superior time and spectrum resolution simultaneously \cite{han2023mre}.
Consequently, these systems often prioritize spectral resolution, potentially degrading performance for various temporal utterances.
At this point, we introduced a multi-resolution encoder (MRE) designed to capture information at different temporal resolutions.
The MRE, a module proposed in Han $et$ $al.$ \cite{han2023mre}, extracts multiple features at various temporal resolutions.
We specifically tailored this module for use in the first-layer module of a raw waveform-based system, naming it multi-resolution feature extractor (MRFE).
Thus, the MRFE is able to consider both temporal and spectral resolution simultaneously,deriving features that maintain robustness across various lengths.

In addition, we propose a new bottleneck named the multi-resolution attention (MRA) block.
The MRA combines the advantages of Res2Dilated block \cite{desplan2019ecapa} and an extended dynamic scaling policy (EDSP) \cite{rawnext}.
The Res2Dilated block gradually builds up the temporal context using dilated convolutional layers and hierarchical residual connection \cite{res2net}.
The EDSP, an extension of the Elastic \cite{wang2019elastic}, trains the network to operate dynamically based on the scale of the data.
These two modules each focus on different aspects of the temporal context: the Res2Dilated block focuses on broader contexts, while the EDSP considers variety of contexts.
Meanwhile, various studies have focused on exploring different temporal resolutions to effectively maximize speaker information from utterances for robustness against utterance length \cite{jung2020msa, liu2022mfa}.
By merging these two modules, the MRA block can consider a wider and more diverse temporal context, thereby enhancing robustness to changes in utterance length.

We finally propose a novel structure called MR-RawNet, which is robust to variations in utterance length by utilizing both the MRFE and the MRA.
Experimental results on VoxCeleb1\&2 datasets \cite{voxceleb1, voxceleb2} demonstrated that MR-RawNet showed superior performance for variable duration utterances in raw waveform-based systems.

\vspace{-0.15cm}
\section{Baseline}

\begin{figure}[t]
  \centering
  \vspace{-0.3cm}
  \includegraphics[width=\linewidth]{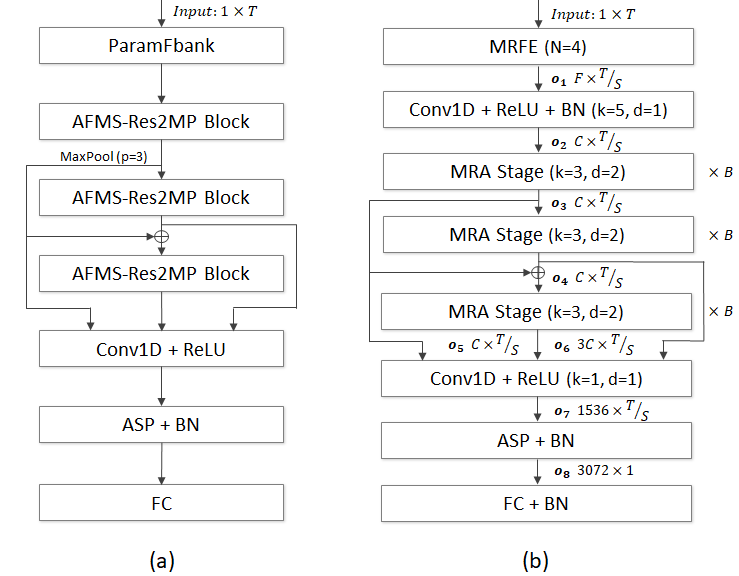}
  \caption{(a): Baseline structure. The kernel size of Conv1D is 1, and $p$ represents max pooling size. (b): MR-RawNet structure. $k$, $d$, $N$, and $B$ denote kernel size, dilation, the number of feature extractors, and the number of MRA blocks.}
  \label{figure:overall_structure}
  \vspace{-0.3cm}
\end{figure}

In a recent study on SV that utilized raw waveforms as input, RawNet3 \cite{rawnet3} achieved remarkable performance. 
Nevertheless, the authors posit that there is still potential for enhancement in RawNet3. 
This section delineates the architecture of RawNet3 and outlines potential areas for improvement. 

RawNet3 is a modified structure of the ECAPA-TDNN \cite{desplan2019ecapa}, which is widely used in SV research. 
As shown in Figure \ref{figure:overall_structure}-(a), RawNet3 is composed of a parameterized analysis filterbank (ParamFbank) \cite{pari2020filter} layer and a Res2Dilated block with $\alpha$-feature map scaling and max pooling (AFMS-Res2MP).
When the one-dimensional raw waveform is fed into RawNet3, it is transformed into a two-dimensional feature map by the ParamFbank layer. 
The ParamFbank learns real-valued parameterised filterbanks as an extension of the SincNet layer \cite{sincnet}.
The feature map, processed through the ParamFbank layer, is then hierarchically passed through three AFMS-Res2MP blocks to capture speaker information at various levels. 
The AFMS-Res2MP is a block that replaces the squeeze-excitation layer in the squeeze-excitation block \cite{seblock} with the $\alpha$-feature map scaling (AFMS) \cite{rawnet2}. 
The extracted speaker information is finally processed into speaker embeddings using convolution, attentive statistics pooling (ASP) \cite{okabe2018asp} with channel- and context-dependent attention values, and linear hierarchy. 

We suggest that there is room for improvement in both the ParamFbank module and the AFMS-Res2MP block within RawNet3. 
Utilizing information from different time scales could allow them to be robust to different lengths of input speech. 

\vspace{-0.15cm}
\section{Proposed methods}
Our proposed architecture enhances the existing RawNet3 to be robust against utterances of various lengths by applying both MRFE and MRA methods.
Figure \ref{figure:overall_structure}-(b) illustrates the overall structure of our proposed system.
The MR-RawNet is a structure that replaces the first convolutional module and bottleneck blocks in the baseline with The MRFE and MRA.
The raw waveform of length $T$ is fed to the MRFE module to generate a time-frequency representation.
The MRFE consists of $N$ feature extractors (FEs) and combines the outputs of the FEs into a feature map $o_1 \in \mathbb{R}^{F \times \frac{T}{S}}$, where $F$ and $S$ refer to the channel and stride size of the MRFE, respectively.
The feature map $o_1$ is then fed into a convolutional layer for input into the bottleneck stages, and its output feature $o_2 \in \mathbb{R}^{C \times \frac{T}{S}}$, where $C$ refer to the channel size of the MRA block, is digested to the bottleneck stage.
Each bottleneck stage consists of $B$ MRA blocks, its outputs $o_3, o_4, o_5 \in \mathbb{R}^{C \times \frac{T}{S}}$ are concatenated into a feature $o_6 \in \mathbb{R}^{3C \times \frac{T}{S}}$, similarly to the ECAPA-TDNN.
The concatenated feature $o_6$ is input into a convolutional layer with a kernel size of 1, followed by a ReLU function.
Then, the output $o_7 \in \mathbb{R}^{1536 \times \frac{T}{S}}$ is aggregated into an utterance-level feature, and the speaker embedding is an output obtained by inputting the utterance-level features $o_8 \in \mathbb{R}^{3072 \times 1}$ into a linear layer.

\subsection{Multi-resolution feature extractor} 

\begin{figure}[t]
  \centering
  \vspace{-0.3cm}
  \includegraphics[width=\linewidth]{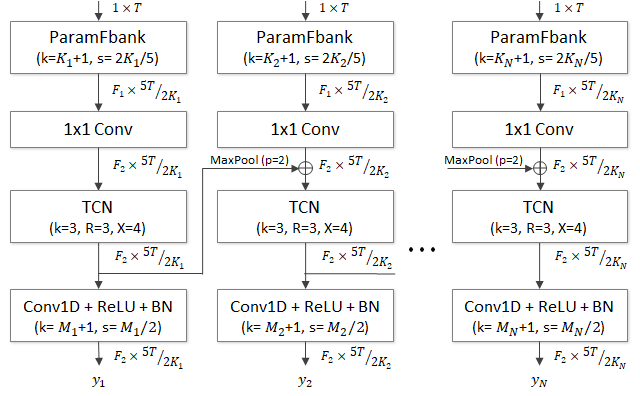}
  \caption{MRFE structure. $k$, $s$, $R$ and $X$ denote kernel size, stride size, the number of repeats, and the number of convolutional blocks in each repeat, respectively.}
  \label{figure:MRFE}
  \vspace{-0.3cm}
\end{figure}

When extracting a time-frequency representation from the raw waveform, longer windows yield superior spectral resolution at the cost of temporal resolution, and the opposite is true for shorter windows.
The first-layer modules of existing raw waveform-based SV systems typically used windows of a fixed frame size.
However, features derived from a fixed frame size cannot concurrently yield excellent time and frequency resolution.
From this perspective, we designed to combine multi-resolution information from the raw waveforms at a low-level to generate a robust time-frequency representation for different utterance lengths.
The MRFE module compresses the information contained in the raw waveform at a low level.
This is similar to extracting handcraft features, but has the advantage of being able to extract features suitable for a specific task using a data-driven method of DNN.

Figure \ref{figure:MRFE} illustrates the structure of the MRFE.
In order to extract discriminative representations from the raw waveform, we used $N$ ParamFbank layers in parallel.
The $i$-th ParamFbank layer (1 $\leq$ $i$ $\leq$ $N$) has a kernel size of $K_{i}$ and a stride size of $\frac{2K_{i}}{5}$, and outputs a time-frequency feature $\in \mathbb{R}^{F_{1} \times \frac{5T}{2K_i}}$.
The feature is then fed to the convolutional layer with kernel size 1 (1$\times$1 Conv) for input to a temporal convolutional network (TCN).
The TCN is a structure proposed to replace recurrent neural network in various sequence modeling tasks, and is used in various fields to extract speech features \cite{lea2016tcn, convtasnet, bai2018tcn}.
The TCN consists of stacked dilated 1-D convolutional blocks, where the dilation factor increases exponentially to ensure a sufficiently large temporal context window.
The 1-D convolutional block is composed of one depthwise convolution between two pointwise convolutions.
The parametric rectified linear unit (PReLU) \cite{prelu} and global layer normalization (gLN) \cite{convtasnet} are used as a nonlinear activation function and a normalization technique in the blocks, respectively.
We used $X$ convolutional blocks with dilation factors $1,2,\cdots,2^{X-1}$ repeated $R$ times.

The $i$-th TCN output $\in \mathbb{R}^{F_{2} \times \frac{5T}{2K_i}}$ passes through a last convolutional layer, which is processed into a time-frequency representation $y_i \in \mathbb{R}^{F_2 \times \frac{T}{S}}$.
The $i$-th last convolutional layer has a kernel size of $M_{i}$ and a stride size of $\frac{M_{i}}{2}$.
We set the feature extractor (FE) stride size $S$ to $\frac{K_{i}\times M_{i}}{5}$ to ensure that outputs of all FEs have the same temporal dimension.
Additionally, the $i$-th output of TCN is also added to the ($i$+1)-th FE using a max pooling with a downsampling factor of 2.
The kernel sizes $K_{i+1}$ and $M_{i+1}$ of the ($i$+1)-th FE are defined as $K_{i+1}=2K_{i}$ and $M_{i+1}=\frac{M_{i}}{2}$, respectively.
Finally, the MRFE outputs a time-frequency representation $\in \mathbb{R}^{F\times \frac{T}{S}}$ ($F$ = $F_{2}\times N$) by concatenating FE outputs to the frequency axis.

\subsection{Multi-resolution attention (MRA) block}

\begin{figure}[t]
  \centering
  \vspace{-0.3cm}
  \includegraphics[width=0.85\linewidth]{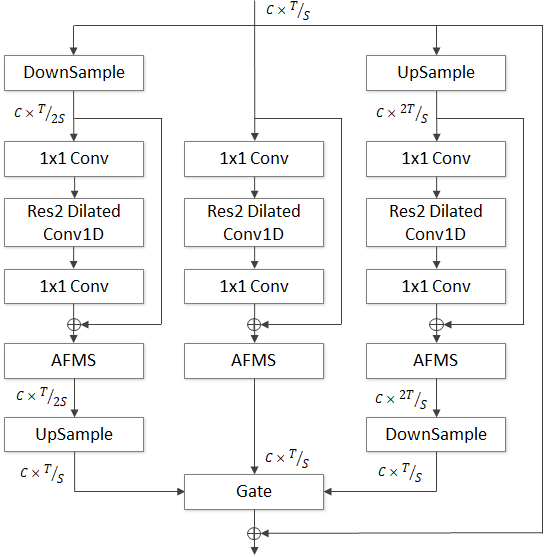}
  \caption{MRA structure. DownSample and UpSample refer to the transposed convolution and average pooling, respectively. The scale dimension of Res2Dilated Conv1D is 4.}
  \label{figure:MRA}
  \vspace{-0.3cm}
\end{figure}

More diverse temporal contexts may need to be considered to create the network that is robust to the length variation of the data \cite{rawnext, wang2019elastic}.
Also, the network performance benefits from a wider temporal context according to the results of \cite{desplan2019ecapa, snyder2019srms}.
Considering this, we created a MRA block based on Res2Dilated block and extened dynamic scaling policy (EDSP) modules.
The Res2Dilated block progressively amasses the temporal context and provides an expansive receptive field.
The EDSP enables the network to function dynamically contingent on data scale.
These two modules each independently focus on a wider and more diverse temporal context.
Therefore, MRA block focuses more on time context to be robust to changes in utterance length.

Figure \ref{figure:MRA} illustrates the structure of the MRA block.
The MRA block receives a feature $\in \mathbb{R}^{C \times \frac{T}{S}}$ as input, and extends the feature resolution range by adding low-resolution and high-resolution paths in parallel from the original.
The low-resolution branch uses a down-sampling function to lower the temporal resolution of the input, and the high-resolution branch uses a up-sampling function to increase the temporal resolution of the input.
We used a 1-D transposed convolution layer and an average pooling layer with kernel size 2 as the down- and up-sampling function respectively.
Then, the resolution-converted inputs are processed through Res2Dilated block with $\alpha$-feature map scaling (AFMS-Res2Block) of the same structure in each branch.
Applying the same structure block at different temporal resolutions means extracting features with receptive fields of different sizes.
Indeed, the branch extension provides the ability to process features with various combinations of receptive fields compared to fixed single-scale branches.
Thereafter, AFMS-Res2Block outputs from the low- and high-resolution paths are then converted to match the original temporal resolution $\frac{T}{S}$ through a sampling function.

We additionally used an attention-based gate module to concentrate on informative components between features output at different temporal resolutions.
The gate focuses on enhancing the expressiveness of multi-resolution paths by modelling channel-specific relationships.
Let $h_t \in \mathbb{R}^{C \times \frac{T}{S}}$ be a feature output from the low-, high-, and original-resolution branches (1 $\leq$ $t$ $\leq$ 3).
Then, a gate module output $o \in \mathbb{R}^{C\times \frac{T}{S}}$ are calculated as follows:
\vspace{-0.2cm}
\begin{equation}
    o = \sum_{t=1}^3{(\alpha_t \times h_t)} 
\end{equation}
where $\alpha_t$ denotes an attention score for a feature output from the low-, high-, and original-resolution branches.
This attention score $\alpha_t$ is calculated using two linear layers ($W_1,b_1$ and $W_2,b_2$) as follows:  
\vspace{-0.2cm}
\begin{equation}
    \alpha_t = \frac{\exp(z_t)}{\sum^3_{i=1}{\exp(z_i)}}, \alpha_t \in \mathbb{R}^{C\times 1}
\end{equation}
\vspace{-0.2cm}
\begin{equation}
    z_t = W_2(\sigma(W_1\rho(h_t)+b_1))+b_2, z_t \in \mathbb{R}^{C\times 1}
\end{equation}
where $\rho(\cdot)$ and $\sigma(\cdot)$ denote an adaptive average pooling and an activation function followed by batch normalization, respectively.
Finally, the MRA block output $\in \mathbb{R}^{C \times \frac{T}{S}}$ is used by adding the output $o$ and the residual, which is the input of the block.

\vspace{-0.15cm}
\section{Experimental setup}

\begin{table*}[t!]
  \caption{Performance comparison of recently proposed speaker verification systems for short utterances. ($^\dag$: our implementation)}
  \label{table:state-of-the-art}
  \centering
  \vspace{-0.3cm}
  \begin{tabular}{llll|cccc}
    \toprule
    \multirow{2}{*}{} & \multirow{2}{*}{\textbf{Input Feature}}  & \multirow{2}{*}{\textbf{Loss Function}} & \multirow{2}{*}{\textbf{Data Augmentation}} & \multicolumn{4}{c}{\textbf{EER(\%) / MinDCF}}  \\
    \cline{5-8}
     &  &  &  & \textbf{Full} & \textbf{5s} & \textbf{2s} & \textbf{1s} \\
    \toprule
    MSEA-FPM \cite{jung2020msa} & MFB-64 & A-Softmax & - & 1.98 / 0.205 & 2.17 & 3.38 & 5.92  \\
    ResNet34-ANF \cite{kei2021anf} & MFB-40 & Softmax+PN &- & 1.91 / 0.221 & 2.04 & 2.88 & 4.49  \\  
    ECAPA-TDNN$^\dag$ \cite{desplan2019ecapa} & MFB-80 & AAM-Softmax & MUSAN+RIR+SpecAug & 0.95 / 0.062 & 0.98 & 1.79 & 3.94  \\
    \hline
    RawNet2 \cite{rawnet2} & Waveform & Softmax & - & 2.43 / 0.236 & 2.64 & 3.88 & 7.24 \\
    RawNeXt \cite{rawnext} & Waveform & AAM-Softmax & MUSAN+RIR & 1.29 / 0.142 & 1.45 & 2.34 & 4.37 \\
    FDN-W-Res2MP \cite{li2023fdnw} & Waveform & AAM-Softmax & MUSAN+RIR & 1.42 / 0.093 & - & - & - \\
    RawNet3 \cite{rawnet3} & Waveform & AAM-Softmax & MUSAN+RIR+Mask+Speed & 0.89 / 0.066 & \textbf{0.90} & 1.81 & 4.35 \\
    MR-RawNet & Waveform & AAM-Softmax & MUSAN+RIR+Speed & \textbf{0.83} / \textbf{0.063} & 0.99 & \textbf{1.61} & \textbf{3.47} \\
    \bottomrule
  \end{tabular}
  \vspace{-0.3cm}
\end{table*}

\subsection{Datasets}
We utilized the VoxCeleb1\&2 \cite{voxceleb1, voxceleb2} datasets to assess our proposed framework. 
The VoxCeleb1 is divided into two subsets: a development set encompassing 148,642 samples from 1,211 speakers and an evaluation set comprising 4,874 samples extracted from 40 speakers.
The VoxCeleb2 development set consists of 1,092,009 utterances obtained from 5,994 speakers. 
During the training phase, we leveraged both the development portions of VoxCeleb1 and VoxCeleb2, whereas, for the evaluation, the VoxCeleb1 test set was employed.
Additionally, VOiCES development set \cite{voices} with 15,904 utterances and 196 speakers was used to conduct out-of-domain evaluation.
For data augmentation techniques, we utilized the MUSAN corpus \cite{musan} and RIR reverberation datasets \cite{rirs}.
The performance of the models was measured using equal error rate (EER) and the minimum detection cost function (MinDCF) with $P_{Target}$=0.05 and $C_{FalseAlarm}$=$C_{Miss}$=1.

\subsection{Configurations}
We constructed a mini-batch with pre-emphasized raw waveforms of either a randomly cropped length of 3 seconds or a random length between 1 and 3 seconds, each chosen with a fifty percent probability.
Evaluation utterances were cut on both sides of the center to measure performance at various lengths.
Adam optimizer \cite{adam} is employed with weight decay of $5e^{-5}$, and the learning rate is scheduled between $5e^{-4}$ and $3e^{-6}$ with a cosine annealing learning rate \cite{sgdr}. 
For speaker identification training, we utilized AAM-softmax \cite{aamsoftmax} with a margin of 0.3 and a scale of 30.
$K_1$ and $M_1$, the kernel sizes of the first extractor in MRFE, were set to 50 and 16, respectively.
We set $K_i$$\times$$M_i$ to 800, which is equivalent to using a window size of 50 ms and a hop size of 10 ms.
Further details are accessible in figures and our code \footnote{https://github.com/kimho1wq/MR-RawNet}.

\vspace{-0.15cm}
\section{Results}
Table \ref{table:state-of-the-art} compares the performance of our proposed framework with recently proposed SV systems for short utterances across various lengths (1s, 2s, and 5s). 
Although RawNet3 did not provide the performance across various utterance lengths, we measured the performance of short utterances using the official model parameters of the RawNet3.
RawNet3 demonstrated comparable performance to other SV systems for short utterances, even though it exhibited superior performance for full-length utterances.
The proposed MR-RawNet showed outstanding performance in various utterances, not only against raw waveform models but also against other models.
Compared to the RawNet3, MR-RawNet displayed a relative error reduction (RER) rate of 20.2\% in 1-second utterances.
These experimental results suggest that a framework capable of exploring information across various time scales can enhance robustness against variable lengths.

We conducted ablation experiments to validate the effects of the proposed methods as shown in Table \ref{table:ablation}. 
These experiments were performed based on the RawNet3 structure, which served as our baseline system.
Experiments \#1, \#2, \#3, \#4 applied only MRFE to the baseline, showing varying performance based on the number of MRFEs ($N$) applied.
Systems that utilized only one or two resolutions fell short when compared to the baseline.
Yet, the application of three or more MRFEs surpassed the baseline performance.
These outcomes suggest that the incorporation of information from various resolutions is beneficial for improving the system's effectiveness with short utterances.
Through the results of \#4, we confirmed that the model could become more resilient to variable lengths by appropriately processing and utilizing information at various resolutions through different blocks.
Experiments \#5 and \#6 apply MRA to the baseline, reflecting the alterations in MRA channels $C$ and blocks $B$. 
For the consistency of parameter quantities, we decreased the number of channels according to the increase in the number of MRA blocks.
Both experiments achieved improved performance compared to the baseline, notably experiment \#6, which set $C$ and $B$ to 256 and 3, respectively.
Experiments \#7 and \#8 represent the combination of experiments \#4 with \#5 and \#6, respectively, and experiment \#8 outperforms other experiments, recorded an RER of 15.6\% for 1-second utterances against the baseline. 
From these results, we believe that MR-RawNet was able to improve the performance for variable duration utterances further by encouraging the model to focus on complementary temporal context through the MRFE and MRA modules.

Table \ref{table:voices} provides a comparison of our system and the baseline model for out-of-domain utterances of various lengths. 
MR-RawNet evaluation was carried out employing System \#8, which showed the best performance under the VoxCeleb1 test condition.
Despite the reduction in the number of parameters compared to RawNet3, MR-RawNet demonstrated improved performance across various utterance durations in the out-of-domain dataset. 
Specifically, MR-RawNet presented an 13.4\% performance improvement over RawNet3 for 1-second utterances, thus proving its effectiveness for short utterances and the generalizability of the system.

\begin{table}[t!]
  \caption{Results of ablation experiments. $N$, $C$, and $B$ denote number of MRFE, channel size of MRA block, and the number of MRA block.}
  \label{table:ablation}
  \centering
  \vspace{-0.3cm}
  \begin{tabular}{lccc|ccc}
    \toprule
    \textbf{} & \multirow{2}{*}{\boldsymbol{$N$}} & \multirow{2}{*}{\boldsymbol{$C$}} & \multirow{2}{*}{\boldsymbol{$B$}} & \multicolumn{3}{c}{\textbf{EER(\%)}} \\
    \cline{5-7}
    & & & & \textbf{Full} & \textbf{2s} & \textbf{1s} \\
    \midrule
    \#0-Baseline  & $\times$ & $\times$ & $\times$ & 1.01 & 1.96 & 4.11 \\
    \hline
    \#1-MRFE  & 1 & $\times$ & $\times$ & 1.16 & 2.17 & 4.68 \\
    \#2-MRFE  & 2 & $\times$ & $\times$ & 1.04 & 1.95 & 4.27 \\
    \#3-MRFE  & 3 & $\times$ & $\times$ & 0.96 & 1.83 & 4.01 \\
    \#4-MRFE  & 4 & $\times$ & $\times$ & 0.93 & 1.79 & 3.98 \\
    \hline
    \#5-MRA & $\times$ & 384 & 1 & 1.03 & 1.93 & 4.08 \\
    \#6-MRA & $\times$ & 256 & 3 & 0.86 & 1.65 & 3.85 \\
    \hline
    \#7-MR-RawNet & 4 & 384 & 1 & 0.92 & 1.77 & 3.69 \\
    \#8-MR-RawNet & 4 & 256 & 3 & \textbf{0.83} & \textbf{1.61} & \textbf{3.47} \\
    \bottomrule
  \end{tabular}
\end{table}

\begin{table}[t!]
  \caption{Results of out-of-domain experiments on the VOiCES development set.}
  \label{table:voices}
  \centering
  \vspace{-0.3cm}
  \begin{tabular}{lc|ccc|c}
    \toprule
    & \multirow{2}{*}{\textbf{\# Params}} & \multicolumn{4}{c}{\textbf{EER(\%)}} \\
    \cline{3-6}
    & & \textbf{5s} &  \textbf{2s} &  \textbf{1s} & \textbf{Avg}\\
    \midrule
    RawNet3 & 16.3M & 2.52 & 6.98 & 13.48 & 7.66 \\
    MR-RawNet & \textbf{15.5M} & 2.52 & \textbf{5.96} & \textbf{11.67} &  \textbf{6.72} \\
    \bottomrule
  \end{tabular}
  \vspace{-0.3cm}
\end{table}

\vspace{-0.1cm}
\section{Conclusion}
We proposed MR-RawNet, which is a novel speaker verification system that is robust to various duration utterances. 
Our system uses MRFE and MRA to improve the performance of variable duration utterances by focusing more on complementary temporal context.
The results on VoxCeleb show that MR-RawNet outperformed other raw waveform-based systems, notably improving performance by 20.2\% in 1-second test compared to RawNet3.
Although this paper focused on comparison with the raw waveform-based models, future research could conduct comparison with models using a variety of input features.

\clearpage

\ifinterspeechfinal
     
\else
     
\fi

\section{Acknowledgement}
This work was supported by Institute of Information \& communications Technology Planning \& Evaluation(IITP) grant funded by the Korea government(MSIT) (No.RS-2023-00263037, Robust deepfake audio detection development against adversarial attacks)

\bibliographystyle{IEEEtran}
\bibliography{mybib}

\end{document}